\documentclass[useAMS,usenatbib]{mn2e}
\usepackage{graphicx}
\usepackage{epstopdf}
\usepackage{amsmath}

\usepackage{times}
\title[]{The influence of large scale magnetic field in the structure of supercritical accretion flow with outflow}
\author[Maryam Ghasemnezhad and Shahram  Abbassi ]{
Maryam Ghasemnezhad $^{1}$\thanks{E-mail:m$_{-}$ghasemnezhad2005@yahoo.com} \thanks{m.ghasemnezhad@uk.ac.ir} and
Shahram Abbassi $^{2,3}$ \thanks{E-mail:abbassi@ipm.ir} \\
$^{1}$Faculty of physics, Shahid Bahonar University of Kerman, Kerman, Iran\\
$^{2}$Department of Physics, School of Sciences, Ferdowsi University of Mashhad, Mashhad, 91775-1436, Iran\\
$^{3}$School of Astronomy, Institute for Studies in
Theoretical Physics and Mathematics, P.O.Box 19395-5531, Tehran, Iran}
\date{}
\begin{document}
\pagerange{\pageref{firstpage}--\pageref{lastpage}} \pubyear{2017}

\maketitle \label{firstpage}
\begin{abstract}

We present the effects of ordered large scale magnetic field on the structure of supercritical accretion flow in the presence of outflow. In the cylindrical coordinates ($r, \varphi,z$), We write the 1.5 dimensional, the steady state ($\frac{\partial}{\partial t}=0$) and axisymmetric ($\frac{\partial}{\partial \varphi}=0$) inflow-outflow equations by using the self similar solutions. Also a model for radiation pressure supported accretion flow threaded by both toroidal and vertical components of magnetic field has been formulated.
 For studying the outflows, we adopt a radius dependent mass accretion rate as $\dot{M}=\dot{M}_{out}{(\frac{r}{r_{out}})^{s+\frac{1}{2}}}$ with $s=\frac{1}{2}$. Also by following the previous works, we have considered the interchange of mass, radial and angular momentum and the energy between inflow and outflow. we have found numerically that two components of magnetic field have the opposite effects on the thickness of the disc and similar effects on the radial and angular velocities of the flow. We have found that the existence of the toroidal component of magnetic field will lead to increasing of radial and azimuthal velocities as well as the relative thickness of the disc,
which is increased . Moreover, the thickness of the disc will decrease when the vertical component of magnetic field becomes
important in magnetized flow.
The solutions indicated that the mass inflow rate, the specific energy of outflow affect strongly on the advection parameter.We have shown that by increasing the two components of magnetic field, the temperature of the accretion flow will decrease significantly. On the other hand we have shown the bolometric luminosity of the slim discs for high values of $\dot{m} (\dot{m}>>1)$  is not sensitive to mass accretion rate and is kept constant ($L \approx 10 L_{E}$).
\end{abstract}

\section{INTRODUCTION}
The study of processes of gas accretion into compact objects started in the 1960s. The standard model, the supercritical accretion model and the radiatively inefficient accretion flow (RIAF) model are three models of accretion flow which have successfully explained energetic phenomena including active galactic nuclei (AGN), ultra-luminous X-ray sources, Gamma-ray bursts and so on.
The optically thick supper Eddington accretion flows (hear after is called Slim discs) belong to the class of cold  accretion flow. These flows proposed by Begelman \& Meier (1982) and also Abramowicz et al. (1988) could describe the cold accretion flows structure in the ultra-luminous X-ray sources and narrow line Seyfert galaxies which cannot be explained by the traditional standard thin disc (Watarai et al. 2001, Mineshige et al. 2000).\\ In contrast to the cold standard model, the slim discs are radiatively inefficient because of both energy advection and photons trapping effects. Consequently, radiation is trapped because of the long photon diffusion time in the vertical direction and advected inward with the accretion flow (Kato et al 2008 for more details). As a result, the radiation pressure is dominant over gas pressure in slim discs (Mineshige et al. 2000). However similar to the standard disc scenario, supercritical accretion flows emit blackbody-like radiation. The standard model of accretion disc breaks down both in the low luminosity regimes (optically thin ADAFs) and high luminosity regimes (optically thick ADAFs or slim discs).\\
Some radiation-hydrodynamic (RHD) and radiation magnetohydrodynamic (RMHD) numerical simulations have shown the existence of outflow in supercritical accretion discs (Okuda et al. 2005, Ohsuga et al 2005, 2009). The radiation-hydrodynamic (RHD) simulations of Ohsuga et al. (2005) have demonstrated the structure of slim discs are composed of the disc region and outflow region above and below the disc. Also Ohsuga et al. (2005) have found that the mass accretion rate is not constant with radius and decreases inward. Blandford \& Begelman (1999) preposed that the radial dependency of the mass accretion rate is as $\dot{M}\propto r^{s}$ with  $0 \leq s\leq 1$ (which the index $s$ is the constant and shows the strong of the wind parameter).\\
 The inward decrease of the inflow accretion rate is related to mass loss in the form of outflow similar to hot accretion flows. Yung et al. (2014) have found that the outflow can be generated at the surface of the disc because of the radiation pressure force and convection instability when the viscose parameter $\alpha$ is large. But for the small viscose parameter the slim discs tend to be convectively stable and the radiation pressure force is the dominant parameter for generation of outflow. The radiatively driven outflow has been extensively considered by many researchers (e.g., Bisnovatyi-Kogan \& Blinnikov 1977; Watarai \& Fukue 1999). \\
 The magnetic field plays a crucial role in the dynamical structure and the
observational properties of accretion disc. Kaburaki (2000), Abbassi et al. (2010 ), Ghasemnezhad et al.(2012), Ghasemnezhad \& Abbassi (2016), Samadi \& Abbassi (2016), Ghasemnezhad \& Abbassi (2016) and Samadi et al. (2014) studied the effect of magnetic field on hot accretion flow (ADAF) in the recent years. But the magnetized slim disc model has been less studied. Ghasemnezhad et al. (2013) have studied the observational properties of the magnetized slim discs without wind parameter in two cases such as LMC X-3 and narrow-line seyfert 1 galaxies.\\
 So the importance and presence of magnetic field in the accretion discs is generally accepted. The magnetic field has several effects such as the formation of the wind/jet (Yuan et al. 2015), the convective stability of accretion flow (Yuan et al. 2012b and Narayan et al. 2012) and the transfer of angular momentum by MRI process in accretion flow (Balbus \& Hawley 1998 )in the dynamical and observational appearance of the discs.\\ 
So the large scale ordered magnetic field is another mechanism to produce outflow in both cold and hot accretion flow. But the magnetic field as the origin of outflow in the hot accretion flow is more important than cold accretion discs.
Three-dimensional Radiation Magnetohydrodynamic (RMHD) simulations by Ohsuga (2009) revealed that the slim disc model supported by radiation pressure which is dominant over the gas and magnetic pressures. The magnetic pressure is two times as large as the gas pressure. This simulation has shown that the toroidal component of magnetic field is dominant in the disc inner region and near the pole the poloidal component of magnetic field in the vertical direction (along the jet axis) is dominant. So it is a valuable work to study the magnetic field on the structure of supercritical accretion flows.\\
The properties of self-similar solutions for the slim disc is similar to ADAFs and have been pointed out by (Fukue 2000). Furthermore the first self-similar solution for a magnetized ADAFs have been presented by Akizuki \& Fukue (2006) which is able to reproduce general properties of slim discs.
 Zahra Zeraatgari et al. (2016) studied the 1.5 dimensional self similar inflow-outflow equations of supercritical accretion flows with outflow in the absence of the magnetic field. They have followed the method of Bu et al. (2009) and Xi \& Yuan (2008) to study the effects of outflow by considering the interchange of mass, radial/angular momentum and energy between inflow and outflow. 
In this paper, we develop Zahra Zeraatgari et al. (2016) solutions by considering the
effects of large scale ordered magnetic field in the 1.5 dimensional inflow-outflow equations.\\   
Hypothesis of the model and relevant equations are developed in
Section 2. Self-similar solutions are presented in Section 3. We
show the result in Section 4.

\section{Basic Equations}
In this paper, we study the effects of large scale magnetic field in the structure of supercritical accretion flows with existent of outflow by using the method presented by Bu et al. (2009) and Xi \& Yuan (2008). In the cylindrical coordinates ($r,\varphi,z$), we write the 1.5 dimensional, the steady state ($\frac{\partial}{\partial t}=0$) and axisymmetric ($\frac{\partial}{\partial \varphi}=0$) inflow-outflow equations using the self-similar approach. Height integrated version of MHD equations (the continuity, the momentum and energy equations) have been adopted, instead of using the two dimensional equations ( e.g. the equations is written in the 1.5 dimension description). Also we use the Newtonian gravity in the radial direction and  also neglect the self-gravity of the discs and the general relativistic effects for simplicity.
 We consider the both $z$ and $\varphi$ components of large scale of magnetic field ($B_{Z},B_{\varphi}$). We do not consider the radial component of magnetic field ($B_{r}$) as we noted in introduction. We investigate the effects of two components of magnetic field to remove angular momentum. Also outflows can be important to remove the angular momentum, mass and energy from the disc. We suppose the $\alpha$-prescription for the turbulent viscosity in the rotating gas
in the accretion flow.  \\
 The MHD equations are follow (Bu et al. 2009) :
 \begin{equation}
 \frac{d \rho}{dt}+\rho\vec{\nabla}.\vec{v}=0,
 \end{equation}
\begin{equation}
\frac{d\vec{v}}{dt}=-\frac{\vec{\nabla} p}{\rho}-\vec{\nabla}\psi+\frac{1}{c}\frac{\vec{J}\times \vec{B}}{\rho}+\frac{\vec{\nabla}\times \vec{T}}{\rho},
\end{equation}
\begin{equation}
\vec{\nabla}\times \vec{B}=\frac{4\pi}{c}\vec{J}.
\end{equation}
\begin{equation}
\vec{\nabla}.\vec{B}=0.
\end{equation}
The induction equation of the magnetic field is:
\begin{equation}
	\frac{\partial \vec{B}}{\partial t}=\vec{\nabla}\times (\vec{v}\times \vec{B}-\frac{4\pi}{c}\eta\vec{J}).
\end{equation}
	The energy equation for the supercritical accretion flows is :
	\begin{equation}
Q	^{+}_{vis}=Q	^{-}_{adv}+Q	^{-}_{rad}.
\end{equation}
where $Q	^{-}_{adv}$ is the advective cooling written as follows:
\begin{equation}
Q	^{-}_{adv}=\rho (\frac{d e}{dt}-\frac{p}{\rho^{2}}\frac{d\rho}{dt})
\end{equation}
In the above equations, $ \rho,\vec{v}, p, e, \vec{B}, \eta, \vec{T}, \psi , Q^{+}_{vis}$ and$ Q	^{-}_{rad}$ have their usual meaning and are the density, velocity, total pressure($p=p_{gas}+p_{radiation}+p_{magnetic}$), gas internal energy $(e=\frac{c^{2}_{s}}{\gamma-1}$ where $\gamma$ is the specific heat ratio and $c^{2}_{s}=\frac{p}{\rho}$ is the sound speed), magnetic field, the magnetic diffusivity, viscous tensor, the Newtonian gravitational potential ($\psi=-GM/r$), viscous heating and radiative cooling respectively. We consider only the azimuthal component of viscous stress tensor as $T=T_{r\varphi}=\rho \nu r \frac{\partial (v_{\varphi}/r)}{\partial r} $ where $\nu$ is the kinematic viscosity. Also the viscous heating is as $Q^{+}_{vis}=T_{r\varphi} r \frac{\partial (v_{\varphi}/r)}{\partial r} $ (Bu et al. 2009, Zahra Zraatgari et al. 2016).
We have followed Bu et al. (2009), Lovelace et al. (1994) and Hirose et al. (2004) assumptions for studying the large scale magnetic field. We consider two components of magnetic field $B_{z}$ and $B_{\varphi}$ in the form of even function of $z$ and odd function of $z$ respectively. Also the vertically gradient of $B_{z}$ is neglected. Hirose et al. (2004) have studied numerically the magnetic field topologies in the different flow regions include of the main disc body, the inner torous and plunging region, the corona envelope and the funnel part. The numerical MHD simulations done by Hirose et al. (2004) have shown that the magnetic field includes two components as: the large scale component and the small scale component (turbulent component). According to these simulations, the effects of turbulence and the differential rotation control the magnetic field geometry in the main body of the disc. The former bends the field line in all directions and the latter removes radial field line into toroidal field line. So the magnetic field configuration of the main disc body is tangled toroidal field and turbulent field. In the plunging and inner regions, the flow changes from turbulence dominated to spiraling inflow. Also in the corona, the magnetic field is purely toroidal without tangles. Although the  large scale poloidal magnetic field only exists in the funnel part. Therefore the toroidal magnetic field governs in the main body of the flow and in its inner parts while the funnel part is governed by a poloidal magnetic fields (which is mostly in the vertical direction $z$). Thus we consider two components of large scale magnetic field $B_{z}$ and $B_{\varphi}$ and we suppose that $B_{r}=0$. We used the $\alpha$-prescription for the turbulent viscosity that consists the all effects of small scale (turbulence) component of magnetic field. In this paper we consider the even $z-$ symmetry field ($ B_{\varphi}(r,z)=- B_{\varphi}(r,-z) $, $B_{z}(r,z)=+ B_{z}(r,-z)$) and $B_{z}$ is in the form of even function of $z$. We can write the magnetic field as follows:
\begin{equation} 
\vec{B}=\vec{B_{p}}(r,z)+\vec{B_{\varphi}}(r,z)\hat{\varphi}
\end{equation}
where $B_{p}$ and $B_{\varphi}$ are the poloidal and toroidal magnetic field components. We can express that $\Delta B_{Z} = \frac{H}{r} (B_{r})_{H}$ from Eq. (4) and by assuming two conditions as $\frac{H}{r} \leq 1$ and $(B_{r})_{H} \leq B_{z}$, it follows that $\Delta B_{Z} << B_{Z}$, that is, we have neglected the variations of $B_{z}$ with $z$ in the disc.
So we have followed these assumptions:
\begin{equation}
B_{r}=0
\end{equation}
\begin{equation}
B_{\varphi}=B_{0}\frac{z}{H}, \rightarrow B_{\varphi,z=H}= -B_{\varphi,z=-H}
\end{equation}
where  $H$ is the half thickness of the accretion disc and  $B_{0}$ to be $B_{0}=B_{\varphi,z=H}$. Also the equation (4) is satisfied by these assumptions. Xie \& Yuan (2008) showed that the accretion flow has two parts: inflow and outflow. They have supposed that the vertical velocity for inflow is $v_{z}=0$ (i.e there is the hydrostatic balance in the vertical direction) and outflow are started from the surface of inflow ($z=H$) with their's own velocity as $v_{z,w}, v_{r,w}, v_{\varphi,w}$. Therefore in the surface of disc there is the discontinuity between outflow and inflow.
 We have written the MHD equations which consist of the coupling between the outflow and inflow in the presence of the large scale magnetic field (Xie \& Yuan 2008, Bu et al. 2009). 
  By integrating of the equation of conservation of mass in the vertical direction and by defining the mass accretion rate as $\dot{M}(r)=-2 \pi r v_{r}\rho $ and the density ratio of outflow and inflow ($\eta_{1}$ ) as $\eta_{1}=\frac{\rho_{w}}{\rho}\cong 0.71$ (Xie \& Yuan 2008), the continuity equation gives: 
\begin{equation}
	\frac{d\dot{M}(r)}{dr}=\eta_{1}4\pi r \rho v_{z,w}.
\end{equation}
The equations of the conservation of radial, angular and vertical momentum are:
\begin{displaymath}
v_{r}\frac{d v_{r}}{dr}+\frac{1}{2\pi r \Sigma}\frac{d \dot{M}(r)}{dr}(v_{r,w}-v_{r})=\frac{v^{2}_{\varphi}}{r}-\frac{GM}{r^{2}}
\end{displaymath}
\begin{equation}
-\frac{1}{\Sigma}\frac{d(\Pi)}{dr}-\frac{1}{4\pi \Sigma}[\frac{d}{dr}(H B^{2}_{z})
+\frac{1}{3}\frac{d}{dr}(H B^{2}_{0})+\frac{2}{3}\frac{B^{2}_{0}}{r}H],
\end{equation}
\begin{displaymath}
\frac{\Sigma v_{r}}{r}\frac{d (rv_{\varphi})}{dr}+\frac{1}{2\pi r}\frac{d\dot{M}(r)}{dr}(v_{\varphi,w}-v_{\varphi})= \frac{1}{r^{2}}\frac{d (r^{3}\nu \Sigma \frac{d\Omega}{dr})}{dr}
 \end{displaymath} 
 \begin{equation}
-\frac{1}{2\pi}B_{Z}B_{0},
 \end{equation}
 \begin{equation}
 \frac{GM}{r^{3}}=(1+\beta_{1})c^{2}_{s}
 \end{equation}
 where $\Sigma \equiv \int \rho dz$ is the vertically integrated density and $\Pi=\int p dz$ is the vertically integrated pressure. By assuming a dominance of the radiation pressure in supercritical accretion discs, we can
write the height-integrated pressure $\Pi=\Sigma c^{2}_{s}=\Pi_{radiation} $. We neglect the gas pressure. All variables are their's own usual meanings like $v_{r}, v_{\varphi}, \Omega$ and $G$ are the radial speed, the rotational speed, the angular speed of inflow and the gravitational constant respectively. Also we use $\alpha$-prescription for the viscosity parameter as ($\nu=\alpha c_{s} H$). Apparently, it is almost impossible to solve the equations directly. We therefore, following Xie \& Yuan (2008), we introduce the parameters $\zeta_{1}$ and $\zeta_{2}$ to evaluate radial and angular velocities of outflow in terms of inflow: $v_{r,w}=\zeta_{1}v_{r},v_{\varphi,w}=\zeta_{2}v_{\varphi}$.

The ratio of magnetic pressure to radiation pressure is assumed as:
\begin{equation}
\beta_{1}=\frac{(B_{\varphi,z=H})^{2}/8\pi}{\rho c^{2}_{s}},
\end{equation}

 \begin{equation}
\beta_{2}=\frac{(B_{z,z=H})^{2}/8\pi}{\rho c^{2}_{s}},
\end{equation}
Three dimensional Radiation Magnetohydrodynamic (RMHD) simulations by Ohsuga (2009) revealed that in the slim disc model the magnetic pressure is smaller than the radiation pressure. So the values of $\beta_{1}$ and $\beta_{2}$ are in the range $\beta_{1}$,$\beta_{2}<1$.
\\
The energy equation:
 \begin{displaymath}
 \frac{\Sigma v_{r}}{\gamma-1}\frac{d c^{2}_{s}}{dr}-2Hc^{2}_{s}v_{r}\frac{d\rho}{dr}+\frac{1}{2\pi r}\frac{d\dot{M}(r)}{dr}(\epsilon_{w}-\epsilon)=
\end{displaymath} 
 \begin{equation}
 \Sigma \alpha c_{s}H r^{2}(\frac{d \Omega}{dr})^{2}-\frac{8\Pi c}{k_{es} \Sigma H}
 \end{equation}
 where $\epsilon_{w}=\zeta_{3} \epsilon$ (Bu et al. 2009) is the specific internal energy of the outflow. The third terms in the left hand side of the Eq. (27) represents that the exchange of internal energy between the inflow and outflow. When $\zeta_{3}>1$, the outflow is extra cooling rate for the inflow and is the extra heating rate for the inflow when $\zeta_{3}<1$.
 \\
 The last term of the energy equation is related to radiative cooling rate which is important in supercritical accretion disc. As we mentioned in introduction, the optically thick ADAFs are supported by radiation pressure. The radiative cooling rate can be written as $Q^{-}_{rad}=\frac{8acT^{4}_{0}}{3\bar{k}\rho H}\cong \frac{8\Pi c}{k_{es}\Sigma H}$ where $a$, $T_{0}$ are the radiation constant and the disc temperature on the equatorial plane. The electron opacity scattering is only considered for the opacity $\bar{k}=k_{es}$ for simplicity.
 \\
 The induction equation (Eq. 5) is the field escaping/creating rate due to magnetic instability
or dynamo effect. For the steady state accretion flow, the advection rate of two components of the magnetic flux will be constant.
  \section{Self-Similar Solutions}
 The self similar method is a dimensional analysis and scaling law and is so useful for solving above equations.
 Following Bu et al. (2009), self-similarity in the
radial direction is supposed:
  \begin{equation}
\Sigma=\Sigma_{0} (\frac{r}{r_{out}})^{s}
\end{equation}
Akizuki \& Fukue (2006) have firstly introduced this type of solution ($\Sigma \propto r^{s}$).
\begin{equation}
v_r(r)=-c_1 \alpha \sqrt{\frac{G M}{r_{out}}}(\frac{r}{r_{out}})^{-\frac{1}{2}}
\end{equation}
\begin{equation}
V_\varphi(r)= r \Omega (r) =c_2 \sqrt{\frac{G M}{r_{out}}}(\frac{r}{r_{out}})^{-\frac{1}{2}}
\end{equation}
\begin{equation}
c_\mathrm{s}^{2}=c_3 \frac{G M}{r_{out}}(\frac{r}{r_{out}})^{-1}
\end{equation}
  The accretion rate decreases inward as:
  \begin{equation}
  \dot{M}(r)=\dot{M}(r_{out})(\frac{r}{r_{out}})^{s+\frac{1}{2}}=2\pi c_{1}\alpha \Sigma_{0}\sqrt{GM} r^{\frac{1}{2}}_{out}(\frac{r}{r_{out}})^{s+\frac{1}{2}}
  \end{equation}
  \begin{equation}
  \dot{M}(r)=2\pi c_{1}\alpha \Sigma_{0}\sqrt{GM} r^{\frac{1}{2}}_{out}(\frac{r}{r_{out}})^{s+\frac{1}{2}}
  \end{equation}
  where $\dot{M}(r_{out})$ and $\Sigma_{0}$ is the mass inflow rate at the outer boundary ($r_{out}$) and the surface density at the outer boundary. Akizuki \& Fukue (2006) have shown that there is some restriction in the parameter $s$. When the parameter $s=-\frac{1}{2}$ there is no mass loss/wind but the self-similar model with outflow could exist only in the case of $s>-\frac{1}{2}$, (we will show that the presence of radiation pressure in the supercritical accretion disc causes that the parameter $s$ was restricted to $s=\frac{1}{2}$). Also Akizuki \& Fukue (2006) have found that for $s=2$, the creation and escape of magnetic field are balanced each other. \\
  By substituting the above self-similar solutions in to the dynamical equations of the
system, we obtain the following system of dimensionless equations, to be solved for $c_{1}$, $c_{2}$ and $c_{3}$:
\begin{displaymath}
-\frac{1}{2}c^{2}_{1}\alpha^{2}-(s+\frac{1}{2})(\zeta_{1}-1)c^{2}_{1}\alpha^{2}=c^{2}_{2}-1-(s-1)c_{3}
\end{displaymath}
\begin{equation}
-\beta_{2}(s-1)c_{3}-\frac{1}{3}\beta_{1}(s-1)c_{3}-\frac{2}{3}\beta_{1}c_{3}
\end{equation}
\begin{displaymath}
\alpha c_{1} c_{2}-2\alpha (s+\frac{1}{2})(\zeta_{2}-1)c_{1}c_{2}=3\alpha c_{3}c_{2}(s+1)\sqrt{1+\beta_{1}}
\end{displaymath}
\begin{equation}
+4\sqrt{\frac{c_{3}\beta_{1}\beta_{2}}{1+\beta_{1}}}
\end{equation}
\begin{equation}
\frac{H}{r}=\sqrt{(1+\beta_{1})c_{3}}
\end{equation}
\begin{displaymath}
\frac{1}{\gamma-1}+(s-1)+\frac{(s+\frac{1}{2})(\zeta_{3}-1)}{\gamma-1}=\frac{9}{4}\frac{c^{2}_{2}}{c_{1}}\sqrt{1+\beta_{1}}
\end{displaymath}
\begin{equation}
-\frac{8}{\dot{m}\sqrt{(1+\beta_{1})c_{3}}}(\frac{r_{out}}{r_{s}})(\frac{r}{r_{out}})^{-s+\frac{1}{2}}
\end{equation}

 For writing the last term of equation (Eq. 27), we use some definitions such as:
 \begin{equation}
  \dot{m}=\frac{\dot{M}_{out}}{\dot{M}_{crit}}
\end{equation}
\begin{equation}
\dot{M}_{out}=2\pi \alpha c_{1} \Sigma_{0} r^{\frac{1}{2}}_{out}
\end{equation}
\begin{equation}
\dot{M}_{crit}=\frac{L_{E}}{c^{2}}=\frac{4\pi GM}{c k_{es}}
\end{equation}
\begin{equation}
r_{s}=\frac{2GM}{c^{2}}
\end{equation}
where $r_{s}$, $L_{E}$ and $c$ are the Schwarzschild radius, the Eddington luminosity and the speed of light respectively.
  The last term in the Eq. (27) which is the radiative cooling rate has the same radial dependency with advection cooling and viscously heating rates when $s=\frac{1}{2}$ (Zahra Zeraatgari et al. 2016). \\
  The second terms on the left hand side of the Eq.(25) is angular momentum transfer between inflow and outflow (the angular momentum taken away from the inflow by the outflow when $\zeta_{2}>1$ or deposited into the inflow by the outflow if $\zeta_{2}<1$). Also in the right hand side of this equation we have the angular momentum transfer by turbulent viscosity and the large scale magnetic field respectively. \\
In the right-hand side of the equation 25 we have two terms which they are represent angular momentum transportation due to turbulent viscosity and large scale magnetic field, respectively. Following Bu et al. 2009 we defining a new parameter $c_{4}$ which is the ratio of these two terms:
\begin{equation}
c_{4}=\frac{3\alpha c_{3}c_{2}(s+1)\sqrt{1+\beta_{1}}}{4 \sqrt{\frac{c_{3}\beta_{1}\beta_{2}}{1+\beta_{1}}}}.
\end{equation}
For given values of $\alpha$, $\dot{m}$, $\zeta_{1}$, $\zeta_{2}$, $\zeta_{3}$, $\beta_{1}$ and $\beta_{2}$ these equations ($c_{1}, c_{2}, c_{3}$ and $c_{4}$) can be worked out numerically. Our results reduce to the results of
Zahra Zeraatgari et al. (2016) without large scale magnetic field. 
  
  \section{Results}
 \subsection{Dynamical structure}
 We are interested to study the effects of large scale magnetic field $(\beta_{1}, \beta_{2})$ on the structure of accretion disc. In all figures, following Zahra Zeraatgari et al. (2016) we adopt the constant values of $\alpha=0.1$, $r_{out}=100 r_{s}$, $s=\frac{1}{2}$, $\gamma=\frac{4}{3}$ and $M=10^{6}M_{\odot}$.
  In six panels of Fig. 1,2, we investigate the effects of large scale magnetic field (azimuthal $(\beta_{1})$ and vertical ($\beta_{2}$) components of magnetic field )
in the radial velocity $(c_{1})$, rotational velocity $(c_{2})$ and the relative thickness $(\frac{H}{r})$ of accretion
flow. Fig. 1,2 shows the coefficients $c_{1}$, $c_{2}$, and $\frac{H}{r}$ in terms of the mass inflow rate $\dot{m}$ at the outer boundary for different values of magnetic parameters $\beta_{1,2}(\beta_{1,2}=0.1,0.2,0.3)$. As we can see easily in two figures, for larger
values of the mass inflow rate $\dot{m}$, the radial velocity and the relative thickness of the disc increase but the rotational velocity decreases. It is in great agreement with Zahra Zeraatgari et al. (2016). Both radial and rotational velocities are sub-keplerian. By increasing mass accretion, more energy will release and consequently the disc becomes thicker. On the other hand the radiation pressure dominated region extends from the inner region of the disc to outer region. Furthermore when the azimuthal magnetic field $B_{\varphi}(\beta_{1})$ becomes stronger in Fig.1, the radial velocity, the rotational velocity and thickness of the inflow increase.
  A magnetized disc must rotate faster and accrete more inward than when there is nonmagnetic field present because of the
effect of magnetic tension force  (the last term in the right hand side of the Eq.(24) is the magnetic stress force ($ -\frac{2}{3}\beta_{1}c_{3}$)). As can be seen, by increasing the toroidal component of magnetic field in the main body of the disc, the disc becomes thicker (see Eq.(26)).
In Fig.2, the vertical component of magnetic field changes from $0.1,0.2$ and $0.3$ Similar to Fig. 1, the radial and azimuthal velocities have an increasing trend with respect to vertical component of magnetic field $\beta_{2}$ while the disc becomes thinner by increasing the vertical component of magnetic field in the high accretion mass. 
Bu et al. (2009) have shown that in the ADAFs model, the large scale magnetic field and outflow cause the reduction of the sound speed of the disc and then the reduction of the thickness of the disc. But we have shown numerically that two components of magnetic field have opposite effect in the thickness of the disc. \\
The effects of large scale magnetic fields components and outflow parameters on the $c_4$ have been plotted in Fig. 3. As we can see in Eq. (32), if $c_{4}<1$, the dominant mechanism in angular momentum transport will be the large scale magnetic field. As the large scale magnetic field in $\beta_{1,2}$ increase, it is obvious that the contributions of turbulent viscosity for transportation of angular momentum decreases which is in a great agreement with Bu et al. (2009). This is because, the large scale magnetic field leads to the inflow gets more angular momentum. In Fig.3 it is clear that when $\beta_{1,2}<0.1$, the effects of large scale magnetic field is almost negligible while for $\beta_{1,2}>0.1$, the parameter $c_{4}$ is below unity and so this yields to decreasing the turbulent viscosity contribution in angular momentum transportation. Also in Fig.3 we have represented the effect of the angular momentum and the specific internal energy of outflow $\zeta_{2,3}$ in the parameter $c_{4}$. We can see from Fig.3 for $\beta_{1,2}<0.1$, $\zeta_{2}$ has a very week effect on the $c_{4}$ while $\zeta_{3}$ has a significant effect on the $c_{4}$. Generally by increasing both $\zeta_{2,3}$, the $c_{4}$ decreases. These results are consistent with solutions of Bu et al. (2009).\\ 
We can calculate the effects of the exchange of the energy and momentum between inflow and outflow and magnetic field on the advection parameter ($f$).  Narayan \& Yi (1994) introduced the advection parameter as $f=\frac{Q^{+}_{vis}-Q^{-}_{rad}}{Q^{+}_{vis}}=\frac{Q^{-}_{adv}}{Q^{+}_{vis}}$. We can write the ratio of the advection cooling to the viscose heating ($f$) as follows:  
  \begin{equation}
  f=\frac{\frac{1}{\gamma-1}+(s-1)+\frac{(s+\frac{1}{2})(\zeta_{3}-1)}{\gamma-1}}{\frac{9}{4}\frac{c^{2}_{2}}{c_{1}}\sqrt{1+\beta_{1}}}
  \end{equation} 

  We have shown the effects of large scale magnetic field $\beta_{1,2}$ and the angular momentum $\zeta_{2}$ and specific energy $\zeta_{3}$ of outflow on the advection parameter $f$ in Fig.4 (the advection parameter $f$ with respect to $\dot{m}$ is plotted). We can see that when the mass inflow rate $\dot{m}$ increases, the advection parameter $f$ increases. When the specific internal energy of outflow ($\zeta_{3}=0.2,0.5$ and $0.9$) increases, the advection parameter also increases (it is simply understood from Eq. (33)). The second term in the surface of the Eq.(33)( $(s+\frac{1}{2})(\zeta_{3}-1)$) is related to interchange of energy between outflow and inflow and for $\zeta_{3}<1$ is negative. So when $\zeta_{3}$ becomes larger, this expression becomes smaller and then the advection parameter increases. As we have mentioned the outflow can play the extra heating role when $\zeta_{3}<1$. On the other hand the advection parameter is independent of the components of magnetic field $\beta_{1,2}$ and the angular momentum of outflow $\zeta_{2}$.
\subsection{The bolometric luminosity}
In the slim disc model, the mass accretion rate and the optical depth are very high. So the radiation generated by accretion disc can
be trapped within the disc and the radiation pressure dominates and sound speed is related to radiation pressure. Slim discs radiate away locally like a blackbody radiation. The emergent local flux $F$ is:
\begin{equation}
F=\sigma T^{4}_{eff}=\frac{1}{2} Q^{-}_{rad}= \frac{16}{3} \frac{\sigma T^{4}_{0}}{\tau}= \frac{4 \Pi c}{k_{es} \Sigma H}
\end{equation}
where $\sigma$ and $\tau(= \frac{1}{2} k_{es} \Sigma)$ are the Stefan Boltzmann constant and the optical depth of the flow respectively.  Factor 2 comes from that the radiation will radiated from both side of the disc. So by using this relation $\frac{\Pi}{\Sigma}=c^{2}_{s}$ and the self-similar solutions we can obtain the emergent local flux and surface temperature of the disc as follows:
\begin{equation}
F(r)=\sigma T^{4}_{eff}=\frac{4c}{k_{es}}\sqrt{\frac{c_{3}}{(1+\beta_{1})}}\frac{GM}{r^{2}}
\end{equation}

 \begin{equation}
 T_{eff}=(\frac{L_{E}}{\pi \sigma})^{\frac{1}{4}}(\frac{c_{3}}{1+\beta_{1}})^{\frac{1}{8}} r^{-\frac{1}{2}}
 \end{equation}
The self-similar solutions for the flux and effective temperature without magnetic field was firstly presented by Watarai \& Fukue (1999). In comparison with their solutions (without magnetic field), in our solutions the flux and effective temperature depend on the toroidal component of magnetic field $\beta_{1}$ explicitly $(\frac{1}{1+\beta_{1}})$ and depend on the vertical component of magnetic field $\beta_{2}$ through $c_{3}$ implicitly.
In Fig.(5), the surface temperature is plotted for different amounts of large scale magnetic field ($\beta_{1,2}$) and mass accretion rates at the outer boundary ($\dot{m}$) in terms of dimensionless radius ($\frac{r}{r_{s}}$). It is clear that the surface temperature decreases as $\frac{r}{r_{s}}$ increases. We can conclude that the larger values of the magnetic field parameters ($\beta_{1,2}$), for a constant radius the temperature of the flow will decrease. A magnetized disc must rotate faster and accrete more inward because of the effect of magnetic stress force (or a centripetal force). As a result, the centrifugal force (both the radiation pressure gradient force $(\propto c_{3})$ and the magnetic pressure gradient force $(\propto c_{3})$) decreases in the presence of large scale magnetic field since the temperature ($c_{3}$) will decreases for the wide range of $\beta_{1,2}$. On the other hand the solution shows that the surface temperature increases as the mass inflow rate at the outer boundary increases that is in agreement with solution presented by Watarai et al. (2000).
 
 It is possible to calculate an analytic expression for $L=L(\dot{m})$. We derive the bolometric luminosity of the slim discs for a typical central black hole with $ M=10^{6} M_{\odot}$ as follows:
 \begin{equation}
 L=2\int_{r_{in}=10 r_{s}}^{r_{out}=100 r_{s}} F(r) 2 \pi r dr
 \end{equation}
The luminosity of the supercritical accretion disc is plotted in Fig. (6) as a function of mass accretion rate for different values of large scale magnetic field $\beta_{1,2}$ and angular momentum carried by the outflow $\zeta_{2}$. The bolometric luminosity of Slim discs increases with increasing mass accretion rate $\dot{m}$.  As a result of photon trapping in the supercritical accretion flows, the maximum luminosity of Slim discs becomes insensitive to the mass accretion rate when $\dot{m}>>1$ ($L/L_{E}\propto \dot{m}^{0} \approx constant$). The disc luminosity is always kept around the Eddington luminosity if the mass accretion rate greatly exceeds unity. We can see easily that the luminosity of the slim discs decreases by adding of two components of magnetic field and the angular momentum carried by outflow and becomes $L \approx 10 L_{E}$ for larger values of $\dot{m}$. These results are in a great agreement with Zahra Zeraatgari et al. (2016) and with simulations performed by Ohsuga et al. (2005) (see Fig. (7) in Ohsuga et al. 2005) and Watarai et al. (2000) (see Fig. (3) in Watarai et al. 2000).
 \begin{figure*}
\centering
\includegraphics[width=138mm]{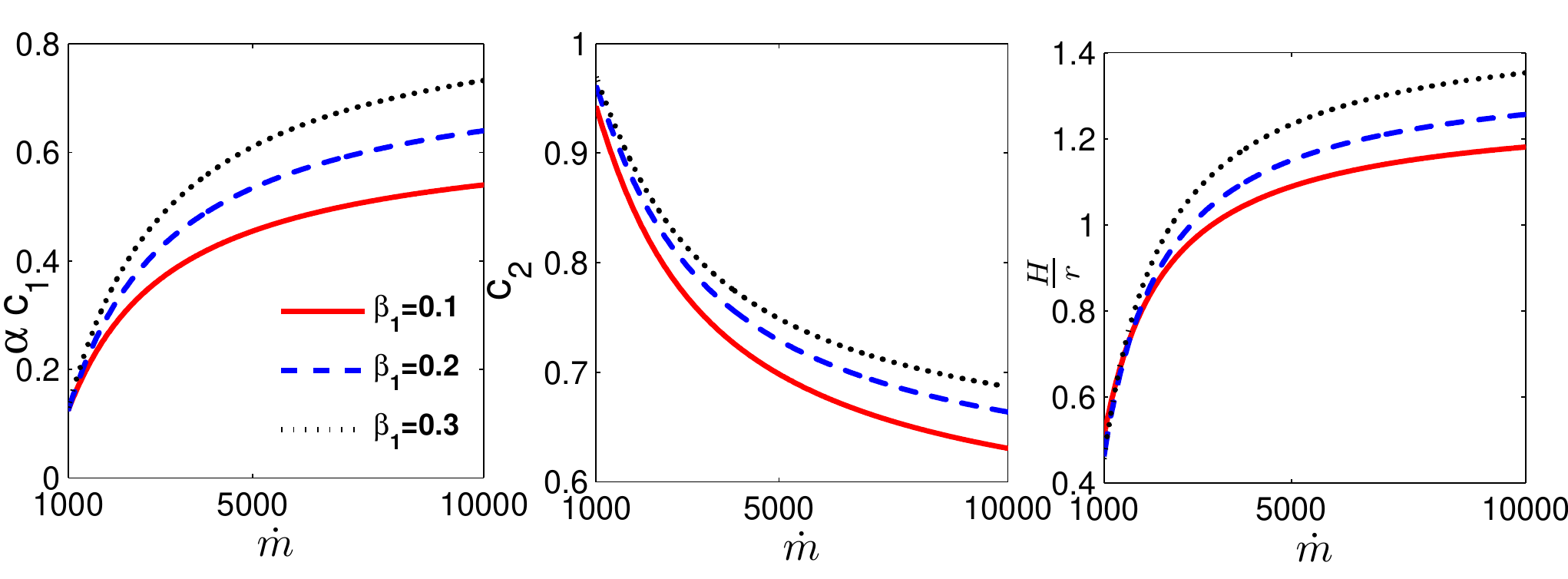}
\caption{ Numerical coefficient $c_1$,$c_2$ and $\frac{H}{r}$ as function of mass inflow rate $\dot{m}$ for several values of $\beta_{1}$ (the toroidal component of magnetic field). For all panels we use $s=0.5$, $\gamma=1.33$, $\alpha=0.1$, $\zeta_{1,2}=0.5$, $r_{out}=100$, $\zeta_{3}=0.2$, $\beta_{2}=0.05$.}
\label{fig1}
\end{figure*}

\begin{figure*}
\centering
\includegraphics[width=138mm]{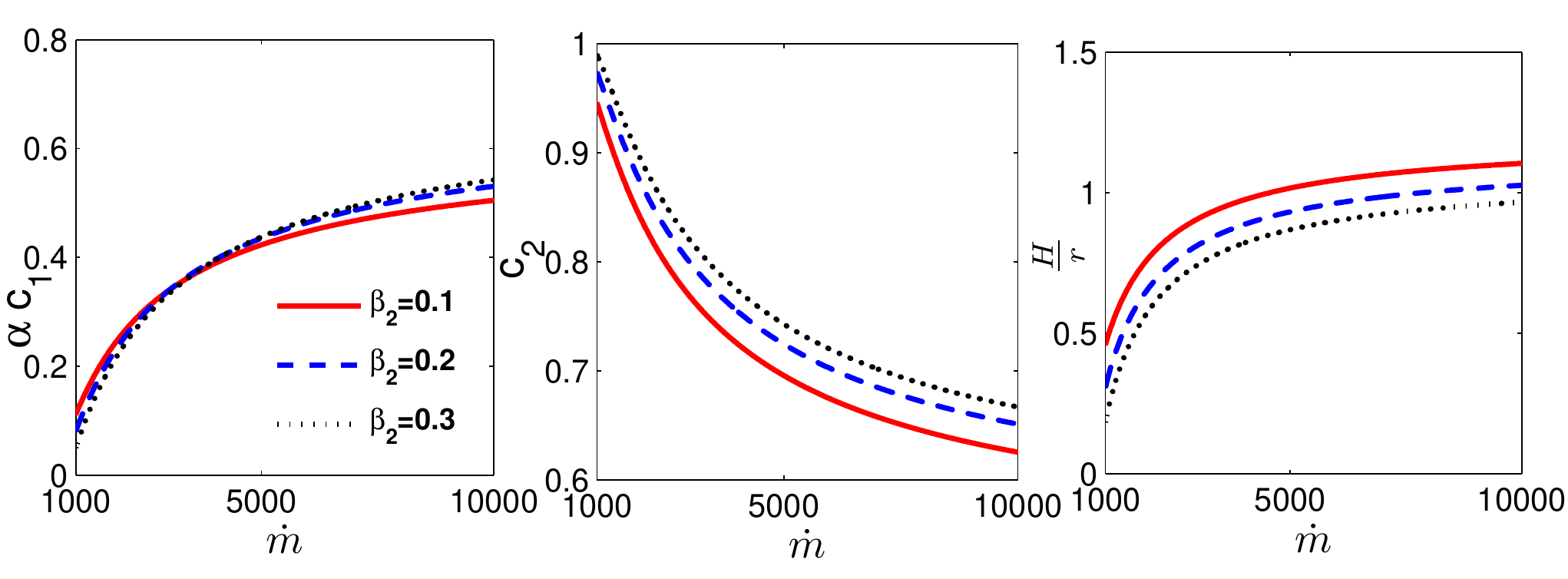}
\caption{ Numerical coefficient $c_1,c_2$ and $\frac{H}{r}$ as function of mass inflow rate $\dot{m}$ for several values of $\beta_{2}$ (the vertical component of magnetic field). For all panels we use
$s=0.5$, $\gamma=1.33$, $\alpha=0.1$, $\zeta_{1,2}=0.5$, $r_{out}=100$, $\zeta_{3}=0.2$, $\beta_{1}=0.05$.}
\label{fig2}
\end{figure*}

  \begin{figure*}
\centering
\includegraphics[width=138mm]{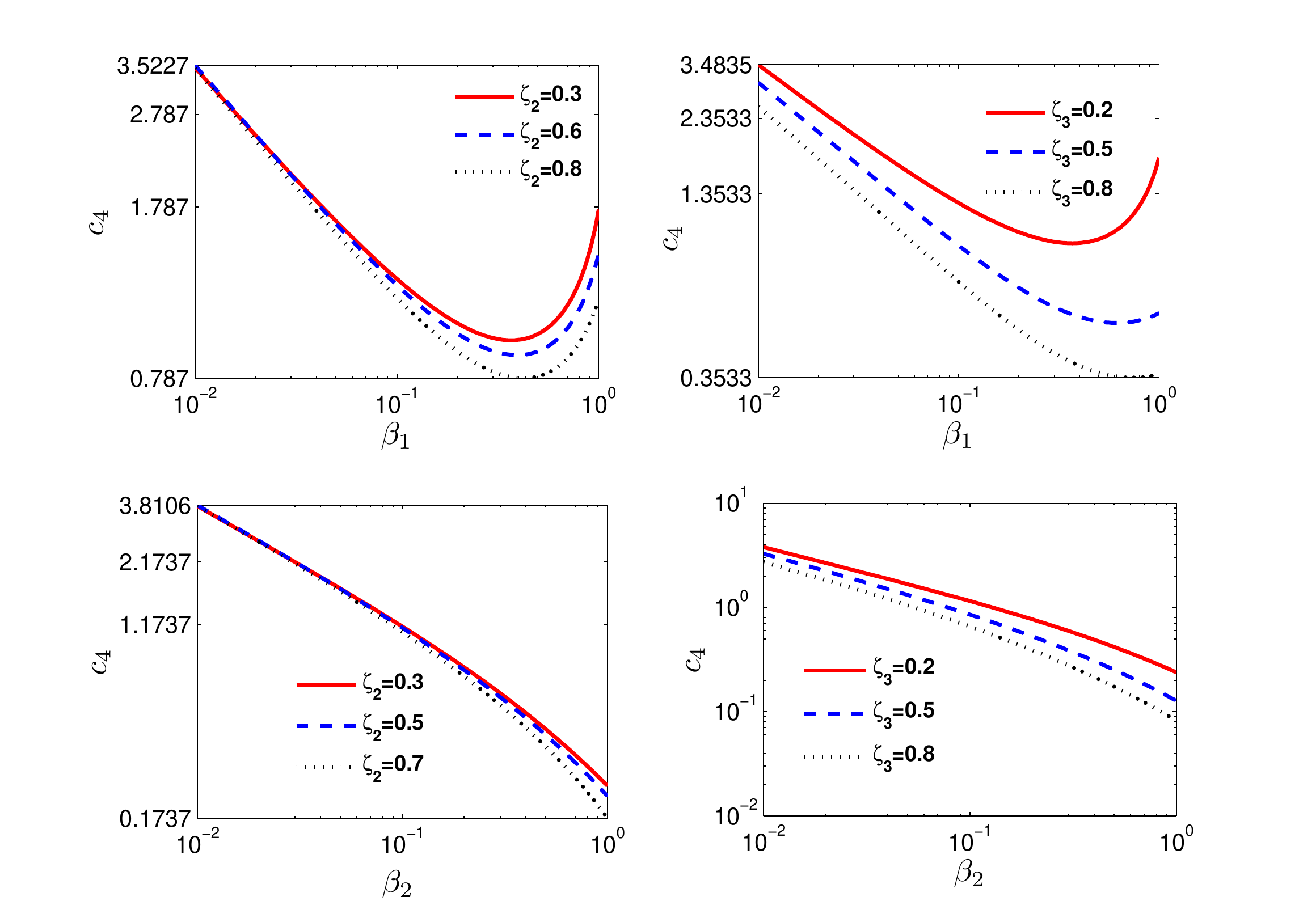}
\caption{  The ratio of angular momentum transport by turbulent viscosity and magnetic field as function of large scale magnetic field $\beta_{1,2}$ for several values of $\zeta_{2,3}$. For all panels we use $s=0.5$, $\gamma=1.33$, $\alpha=0.1$, $\zeta_{1}=0.5$, $r_{out}=100$, $\dot{m}=3000$. In upper and bottom pannel we use $\beta_{2}=0.05$ and $\beta_{1}=0.05$ respectively.}
\label{fig3}
\end{figure*}

  \begin{figure*}
\centering
\includegraphics[width=138mm]{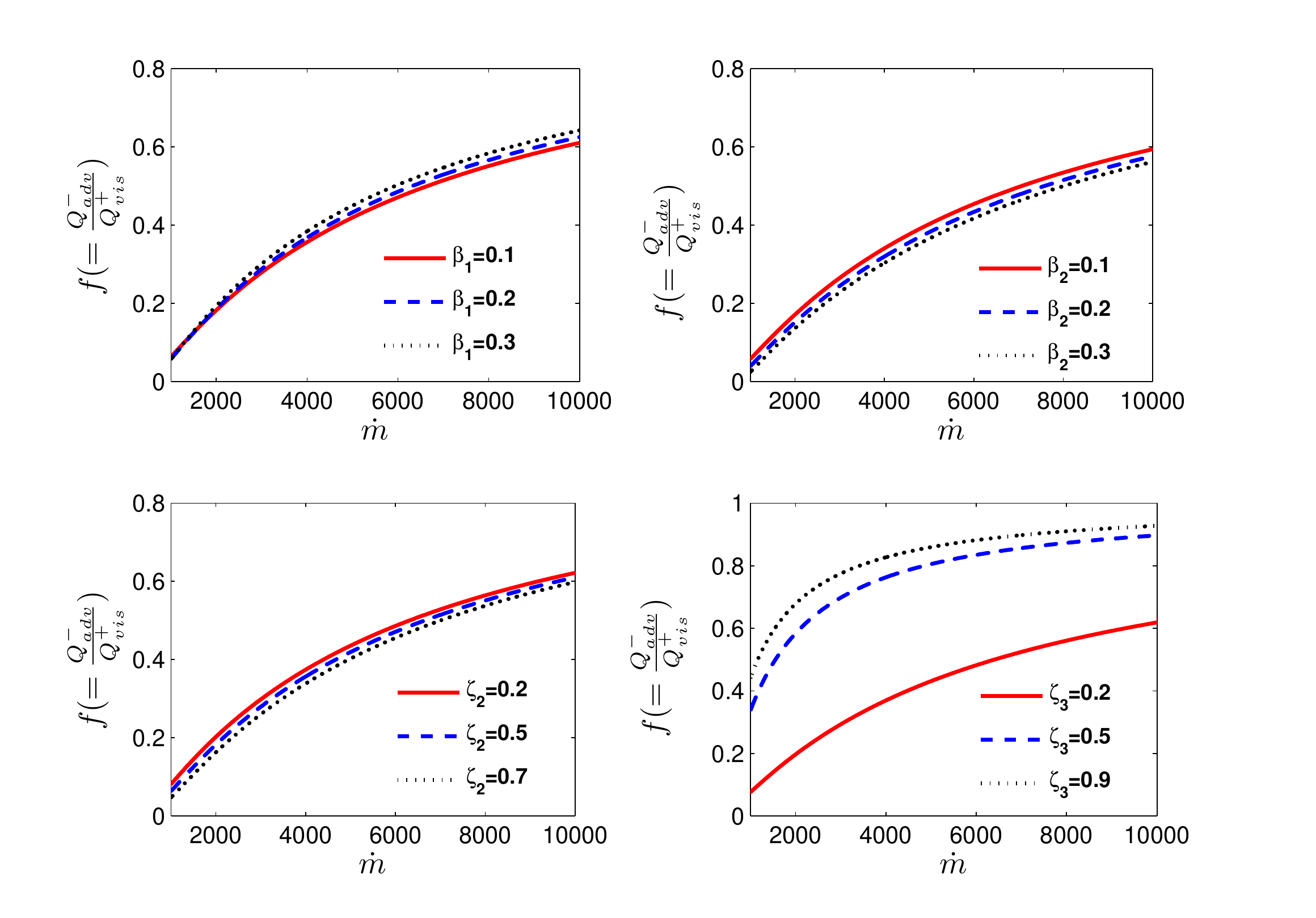}
\caption{The advection parameter $f$ as function of mass inflow rate $\dot{m}$ for several values of $\beta_{1,2}$ and $\zeta_{2,3}$. For all panels we use $s=0.5$, $\gamma=1.33$, $\alpha=0.1$, $\zeta_{1}=0.5$, $r_{out}=100$.}
\label{fig4}
\end{figure*}

  \begin{figure*}
\centering
\includegraphics[width=138mm]{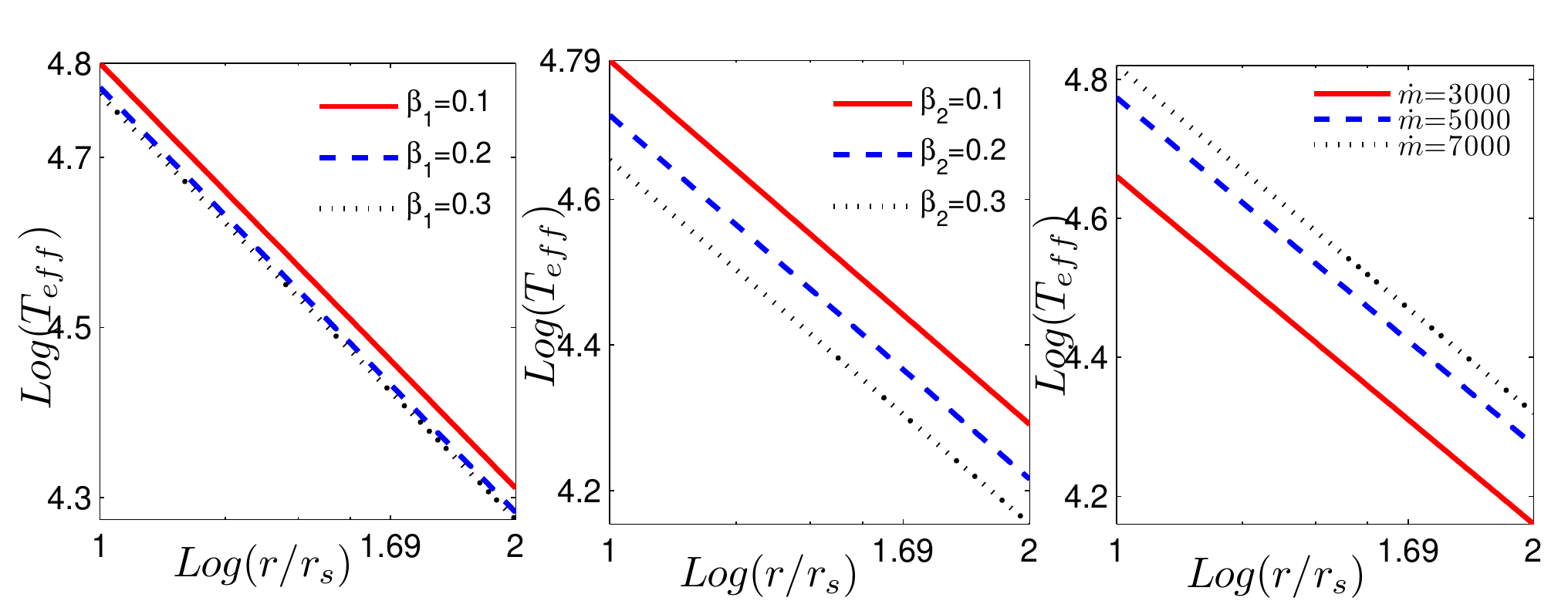}
\caption{The effective temperature as a function of dimensionless radius $r/r_{s}$ for several values of $\beta_{1,2}$ and $\dot{m}$. For all panels we use $s=0.5$, $\gamma=1.33$, $\alpha=0.1$, $\zeta_{1,2}=0.5$, $\zeta_{3}=0.2$, $r_{out}=100$. We set $\beta_{2}=0.05$ for the left pannel, $\beta_{1}=0.05$ for the middle pannel and $\beta_{1}=0.3, \beta_{2}=0.05$ for the right pannel.}
\label{fig5}
\end{figure*}

\begin{figure*}
\centering
\includegraphics[width=138mm]{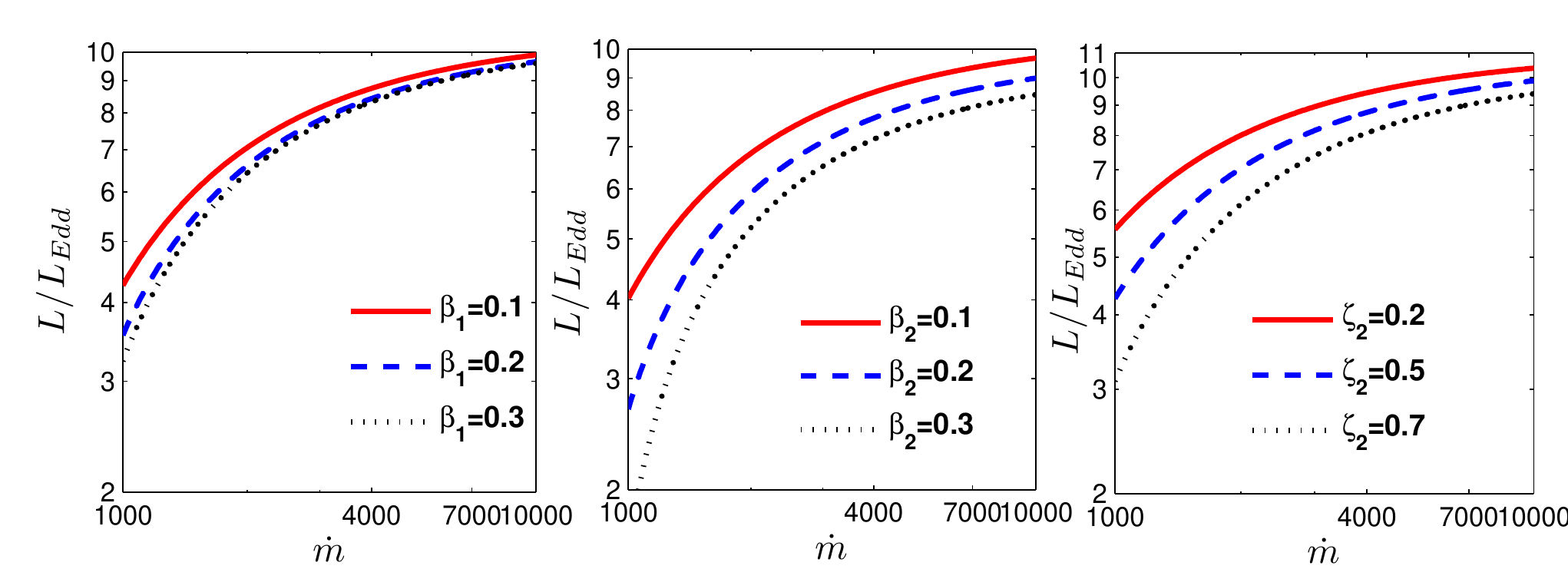}
\caption{Bolometric luminosity as function of mass inflow rate $\dot{m}$ for several values of $\beta_{1,2}$ and $\zeta_{2}$. For all panels we use $s=0.5$, $\gamma=1.33$, $\alpha=0.1$, $\zeta_{1}=0.5$, $\zeta_{3}=0.2$ $r_{out}=100$. We set $\beta_{2}=0.05,\zeta_{2}=0.5$ for the left pannel and $\beta_{1}=0.05, \zeta_{2}=0.5$ for the middle pannel and $\beta_{1}=0.1, \beta_{2}=0.05$ for the right pannel.}
\label{fig6}
\end{figure*}

 \section{Summary and Conclusion}
 In this paper, we have investigated the effects of large scale magnetic field in supercritical accretion flow in the presence of outflow. It was assumed that
outflow contributes to loss of mass, angular momentum,
and thermal energy from accretion discs. 
By using the method presented by Bu et al. (2009) and Xi \& Yuan (2008) we consider the discontinuity between mass, momentum (radial and azimuthal components of momentum) and energy outflow and inflow. Also simulations have indicated that the large scale ordered magnetic field exists in the accretion disc and can produce outflow in both cold and hot accretion flow. RMHD Simulations have shown that the azimuthal component of magnetic field is important in the main body of the disc and the vertical component of magnetic field is dominant near the pole.\\  
 We used the self-similar method for solving the 1.5 dimensional, the stady state ($\frac{\partial}{\partial t}=0$) and axisymmetric ($\frac{\partial}{\partial \varphi}=0$) inflow-outflow equations.  The self-similar solutions are very useful to improve our understanding of the physics of the accretion discs around black hole. 
Also we ignore the relativistic effects and self-gravity of the disc and we use Newtonian
gravity in the radial direction. 
We consider the effect of $B_{\varphi,z}$
on the dynamics of disc by following the paper of Zahra Zeraatgari et al. (2016) in the
presence of the effect of wind. Our results reproduce their solutions 
when the effect of large order magnetic field is neglected.
The disc accrete more and rotates fast in the presence
of large scale magnetic field. Also we have shown that two components of magnetic field have the opposite effects on the thickness and the sound speed of the disc. We find that the strong magnetic field causes the turbulence viscosity contribution in angular momentum transfer decreases. The solutions indicated that the mass inflow rate and the specific energy of outflow have strong effect on the advection parameter.  The outflow is the extra heating rate for the inflow when $\zeta_{3}<1$. We find that by increasing the energy of outflow, the advection parameter increases.\\
Furthermore we have examined how two components of magnetic field affect on the luminosity and the surface temperature of the supercritical accretion disc. As we have mentioned in the previous section, the effective temperature, as well as the bolometric luminosity has a decreasing behavior in respect to two components of magnetic field.\\
 In addition, increasing mass accretion rate makes the temperature and luminosity of disc increases while for high mass accretion rate $\dot{m}>>1$, the luminosity of the disc ($L$) is insensitive to mass accretion rate. We have shown that the maximum luminosity is kept constant and $L \approx 10 L_{E}$ that is in a great agreement with calculations done by Watarai et al. (2000) and Ohsuga et al. (2005).
 Although that we have made some simplification and some limitations in order to solve equations numerically, our results show that ordered large scale magnetic field and outflow can really change structure of supercritical accretion flow which means in any realistic model these factors should take into account. So self-similar solutions could interpret well the numerical simulations and observational evidences.
 
\section{ACKNOWLEDGEMENTS}
SA acknowledges support from department of astronomy, Cornell university for their hospitality during a visit which a part of this work has been done there. We also appreciate the referee for his/her thoughtful and constructive comments on an early version of the paper.

\end{document}